\documentclass[reprint,amsmath,amssymb,superscriptaddress,showpacs,twocolumn]{revtex4-1}
\usepackage{graphicx}
\usepackage{dcolumn}
\usepackage{bm}
\usepackage{amsmath}
\usepackage{amssymb,amsthm}
\usepackage{color}
\usepackage{epstopdf} 
\usepackage{subfig}

\usepackage{epstopdf}
\usepackage{mathbbol}

\usepackage{xcolor,soul} 

\begin{document}

\title{Spin Glass Theory of Interacting Metabolic Networks}

\author{Jorge Fernandez-de-Cossio-Diaz}
\email{j.cossio.diaz@gmail.com}
\affiliation{Systems Biology Department, Center of Molecular Immunology, Havana, Cuba}
\affiliation{Group of Complex Systems and Statistical Physics, Department of Theoretical Physics, University of Havana, Physics Faculty, Cuba}
\author{Roberto Mulet}
\email{mulet@fisica.uh.cu}
\affiliation{Group of Complex Systems and Statistical Physics, Department of Theoretical Physics, University of Havana, Physics Faculty, Cuba}
\affiliation{Italian Institute for Genomic Medicine, IIGM, Torino, Italia}

\date{\today}


\clearpage

\begin{abstract}
  We cast the metabolism of interacting cells within a statistical mechanics framework considering both, the actual phenotypic capacities of each cell and its interaction with its neighbors. Reaction fluxes will be the components of high-dimensional spin vectors, whose values will be constrained by the stochiometry and the energy requirements of the metabolism. Within this picture, finding the phenotypic states of the population turns out to be equivalent to searching for the equilibrium states of a disordered spin model.
  We provide a general solution of this problem for arbitrary metabolic networks and interactions.
  We apply this solution to a simplified model of metabolism and to a complex metabolic network, the central core of the \emph{E. coli}, and demonstrate that the combination of selective pressure and interactions define a complex phenotypic space.
  Cells may specialize in producing or consuming metabolites complementing each other at the population level and this is described by an equilibrium phase space with multiple minima, like in a spin-glass model.


\end{abstract}

\maketitle
\clearpage

Cellular metabolism is defined as the network of chemical reactions that transforms raw materials from the environment into useful products to the cell.
It provides cells with the energy and the building blocks demanded for most biological functions \cite{alberts2015cell}.
Although each type of cell has its own metabolic network \cite{obrien2015c}, many reactions and pathways are conserved across species, and their specific role in metabolism is well characterized.
Prominent examples are glycolysis, the pentose-phosphate pathway and oxidative phosphorylation, which form the backbone of the network of metabolic reactions of most cells.

The recent annotation of genome scale metabolic networks \cite{palsson2015book} lead the study of metabolism to a new level of abstraction and computational difficulty, opening the doors to its quantitative understanding beyond the independent pathway approximation.
In this context, algebraic approaches \cite{schuster1994efm,wortel2014efm}, Linear Programming \cite{palsson2006book} and statistical analysis \cite{wiback2004mc,braunstein2017nc,cossio2017ploscb,cossio2016bp,mori2016cafba,font2012weighted} have become standard tools of the community.

Much less understood is the role of metabolism in the interaction between nearby cells.
The importance of these interactions has been known for many years since they drive the appearance of several diseases \cite{gatenby1996reaction,magistretti2018brain,fischbach2011eating,beckermann2017dysfunctional}.
More generally, cells compete for nutrients, and at the same time they produce by-products that could be either toxic \cite{toussaint2016jb}, alternative sources of energy \cite{faubert2017lac,hui2017n}, or alter the environment \cite{gatenby2008micro}.
This results necessarily in direct or indirect relations between the metabolism of different cells \cite{campbell2015elife}.
However, most of the theoretical work on interacting metabolic networks has been characterized by i) the study of competing/cooperating populations after drastic simplifications of their metabolism \cite{cossio2017micro,smith1995chemostat,wortel2016scirep,mehta2014pnas,monasson2017macarthur}, and ii) the study of the interaction between two individual cells with identical or nearly identical complex metabolism \cite{cossio2016bp,braunstein2017nc,cossio2019maxent,cossio2017ploscb,martino2018natcomm}.
When populations of cells are described by complex metabolic networks, interacting between them or with the environment, simulations have been the main research tool \cite{zelezniak2015pnas,swain2017nc,pfeiffer2001sci,beardmore2011flat}.
An analytic approach to deal with the {\em interaction of complex cells} is currently absent from the literature.

The main purpose of this paper is to build this theory developing an analogy with the physics of classical disordered systems.
We treat the reaction rates of the metabolism in each cell as the components of a continuous spin vector $\vec v$ bounded by thermodynamic constrains $\mathrm{lb}_r \le v_r \le \mathrm{ub}_r$ for each reaction $r$ and representing kinetic limitations of the fluxes in the physiological context of the cell, such as irreversibility. These fluxes must also satisfy the stoichiometric relations of the metabolism in stationary state, \cite{palsson2015book}, $\mathbb{S} \vec v = \vec b$, where $\mathbb S$ is the stochiometric matrix and $\vec{b}$ represents the exchange fluxes with the environment. These constraints define a polytope that contains the possible metabolic states of the cells.

In the absence of interactions it is often considered that each cell $i$ tries to maximize its own utility function, $-E_i = \sum_r h_r v_{ri}$, where the choice of $h_r$ defines the utility function of the cell, and the minus sign accommodates the physical convention that energy is minimized.
Usually, the $h_r$ stand for the growth rate \cite{feist2010com}, ATP production, or maximization of the rate of synthesis of a particular product of interest \cite{varma1993fbaecoli}, etc.
The maximization of this utility function subject to the stochiometric constraints constitutes an optimization problem easily solved by linear programming techniques \cite{vanderbei2014lp}.
However, only in specific situations, such as simple bacteria growing in rich media, cells should be studied as if they optimize an individual utility function.
Cells in a community such as a tissue or a culture share and exchange metabolic products from and with the help of the environment.
In this case, a proper description of the whole system must consider a modified utility function that takes explicitly these interactions into account.

\section*{Interacting cells}

In practice we assume that a collection of cells interacts through a term of the form: $V_{ij} = -\sum_r J_{rij} \phi_{ri} \phi_{rj}$, where the sum over $r$ is done over a set of exchange reactions. If $J_{rij}>0$, a state where both cells $i$ and $j$ carry a similar flux on reaction $r$ is favored.
For example, if reaction $r$ represents consumption of a nutrient such as glucose, then there is an evolutionary pressure favoring \emph{competition} between cells $i$ and $j$ for this nutrient.
On the contrary if $J_{rij}<0$, cells $i$ and $j$ tend to carry opposite fluxes on reaction $r$. If reaction $r$ represents an exchange of a byproduct, such as lactate, it means then that cell $i$ tends to produce lactate and cell $j$ to consume it (or \emph{vice versa}), establishing a so called lactate shuttle \cite{cossio2017micro,brooks1986lactate,pellerin1998evidence,brooks2000intra}).  In this case we say that cells \emph{cooperate}.

Combining the utility functions of individual cells with the terms arising from these pairwise interactions gives rise to the following energy function for the population of cells,
\begin{equation}
H(\{v\}) = -\sum_{r,i} h_i v_{ri} - \sum_{i<j} \sum_r J_{rij} v_{ri} v_{rj}
\end{equation}
which favors states where individual cells attempt to maximize their own utility functions but also try to form favorable interactions with neighboring cells.
Ideally we seek a configuration of cells that minimizes $H(\{v\})$.

In our approach cells are not necessarily identical,  stochastic fluctuations in the synthesis of proteins \cite{kiviet2014singlemetabolism}, evolutionary processes \cite{wortel2016scirep} or local fluctuations in the concentration of raw materials \cite{delvigne2014tb} forces the cell to explore different metabolic states and not only ``optimal'' ones.
We assume that these metabolic states are distributed according to a Boltzmann distribution:
\begin{equation}
  P(\{\vec v\}) = \frac{1}{Z[J]} \exp\left(-\beta H(\{v\})\right)
\end{equation}
provided that the constrains $\mathrm{lb}_r \le v_{ri} \le \mathrm{ub}_r$ and $\mathbf S\vec v=\vec b$ are satisfied, where $Z$ is a normalization constant known in the physics literature as the partition function of the problem, and the parameter $\beta$ quantifies the strength of the stochastic fluctuations that drive cells away from the preferred configurations.
In absence of interactions $\beta$ has been interpreted as an equilibration time-scale in the dynamics of a logistic growth model \cite{martino2018natcomm}.
In a more general setting, $\beta$ was also connected to the mutation rate (or more generally the rate of cell differentiation) in simple dynamic evolutionary models \cite{cossio2019maxent}.
In general, when $\beta \rightarrow 0$, the exponential term is irrelevant and the properties of the problem are defined by the solution space fixed by the stochiometric constraints (a polytope) \cite{braunstein2017nc,cossio2016bp,dyer1988polytope,martino2015mc}.
In the opposite limit, when $\beta \rightarrow \infty$, and in absence of interactions, cells minimize their own $E_i$ subject to the stochiometric constraints. This solution, known in the literature as flux-balance analysis (FBA) \cite{palsson2015book}, lies in a vertex of the polytope and can be found by efficient linear programming techniques \cite{vanderbei2014lp}.



This way to write the problem allows to explore the solution space of {\it interacting} metabolic networks using precise mathematical terms. In addition, this analogy between reaction fluxes and continuous spin variables enlarges the family of disordered systems and makes possible the use of the whole arsenal of concepts, techniques, and approximations already developed in their study in the analysis of metabolic interacting cells \cite{mezard1987book}.

\section*{Mean Field Solution}

To model an heterogeneous population we assume that the couplings $J_{rij}$ are drawn from a normal distribution with mean $J_r$ and variance $\Delta_r$.
For simplicity we take the $h_{ri}=h_r$ as fixed values to be specified below.
The physics of the problem is summarized by a free energy density, defined by $f[J] = -(1/N)\ln Z[J]$ that  must be averaged over the disorder of the couplings.
We denote this operation by an overbar, $f = \overline{f[J]} = -(1/N)\overline{\log Z[J]}$.
To perform the average of the $\log$, we can use the replica trick \cite{mezard1987book}, $\overline{\log Z} = \lim_{n\rightarrow0}\frac{1}{n}\log\overline{Z^n}$
(Details are provided in the Supplementary Material).
We quote here only the final result:
\begin{equation}
\begin{aligned}
  \bar{f} & = - \sum_{\alpha, r} \frac{J_r}{2} m_{\alpha r}^2 -
  \sum_{\alpha < \beta} \sum_r \frac{\Delta_r^2}{2} q_{\alpha \beta r}^2 -
  \sum_{\alpha, r} \frac{\Delta_r^2}{4} \zeta_{\alpha r}^2 \\
   &+ \, \ln \mathrm{Tr} \exp \left(-H_\mathrm{eff}(\{v\})\right)
\end{aligned}
\label{eq:freeenergy}
\end{equation}
where the trace is over all the replicated flux variables $v_r^{\alpha}$, respecting the stoichiometric constrains in each replica, $\alpha=1,\dots,n$. The effective Hamiltonian coupling replicas is given by:
\begin{equation}
\begin{aligned}
H_{\mathrm{eff}} &= - \sum_{r,\alpha} h_r v_r^{\alpha} - \sum_{r,\alpha}
J_r v_r^{\alpha} m_r^\alpha \\
&- \sum_r \sum_{\alpha < \beta} \Delta_r^2
v_r^{\alpha} v_r^{\beta} q_r^{\alpha\beta} - \sum_{r,\alpha}
\frac{\Delta_r^2}{2} (v_r^\alpha)^2 \zeta_r^\alpha
\end{aligned}
\end{equation}
\noindent and the parameters $m_r^\alpha, q_r^{\alpha\beta},\zeta_r^\alpha$ must be chosen to extremize this expression.
Differentiating gives the following set of coupled equations:
\begin{equation}
m_r^\alpha = \langle v_r^{\alpha} \rangle, \quad q_r^{\alpha\beta} = \langle v_r^{\alpha} v_r^{\beta} \rangle, \quad \zeta_{\alpha r} =
\langle (v_r^{\alpha})^2 \rangle
\end{equation}
where $\langle\dots\rangle$ denotes an average with weight $\exp \left(-H_\mathrm{eff}(\{v\})\right)$.


A crucial step is to propose an \emph{ansatz} to the form of the parameters $m_r^\alpha, q_r^{\alpha\beta},\zeta_r^\alpha$ that solve these equations.
Since the replicas are indistinguishable, a natural assumption is that the extremization respects this symmetry, and therefore that $m_r^\alpha = m_r$, $q_r^{\alpha\beta}=q_r$, $\zeta_r^\alpha = \zeta_r$ are independent of the replica index.
This is known as the \emph{replica symmetric} (RS) ansatz, which we assume throughout the rest of the paper.  In short, $q_r$ quantifies the metabolic fluctuations between populations with different realizations of the disorder and $\zeta_r$ the fluctuations in the fluxes between individuals of a population with the same disorder. 

\begin{figure}
\centerline{\includegraphics[width=0.3\textwidth]{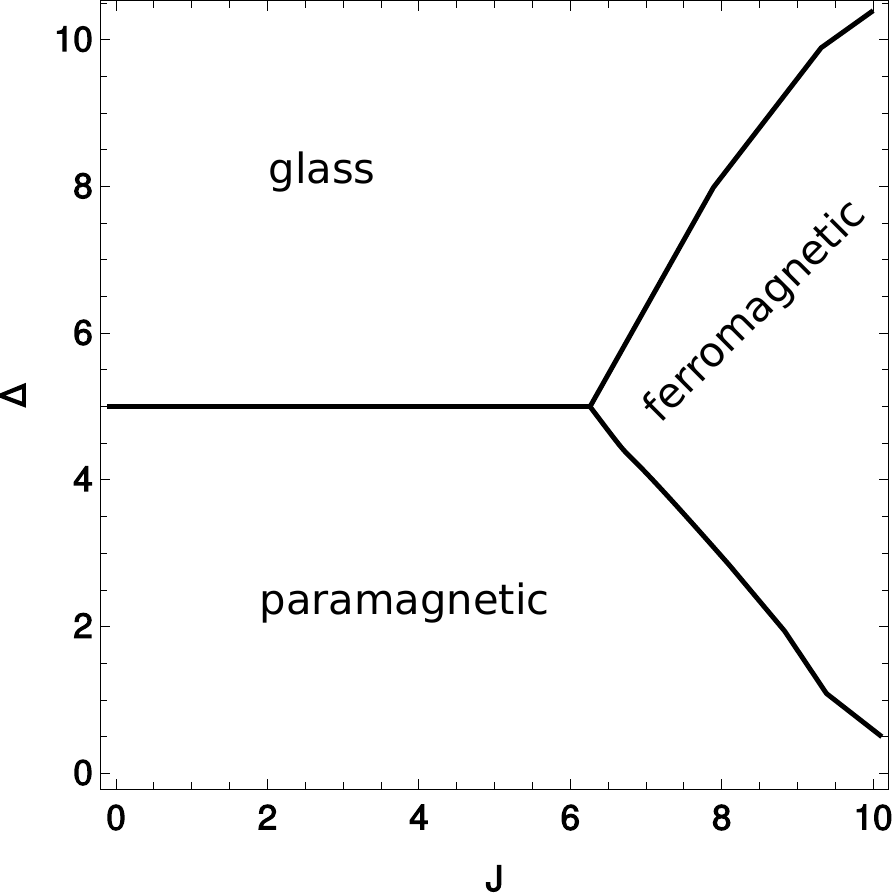}}
\caption{
\label{fig:sgphases} The mean-field model exhibits three phases phases, depending on the disorder in the couplings $\Delta$ and their mean value $J$. The \emph{ferromagnetic} states have non-zero average exchange flux ($v_3$). The \emph{paramagnetic} states have zero average export flux. The \emph{spin glass} states exhibit macroscopic sample-to-sample fluctuations in the average export flux.}
\end{figure}

To first explore this formalism in a manageable model, we considered a simplified metabolic network consisting of three reactions with fluxes $v_i,i=1,2,3$ satisfying the constrains $v_1+v_2+v_3=0$, $-1\le v_1\le0$, and $0\le v_2\le1$. 
The fluxes $v_i,i=1,2,3$ can be interpreted as glucose consumption ($v_1$), respiration ($v_2$), and lactate secretion/consumption ($v_3$), respectively. It has two metabolic modes: the respiration mode with $v_3\le0$ and the fermentation mode with $v_3\ge0$.
To model the contribution of these modes to energy production in the cell, we set $h_2 = -h_1 = h = 1$, $h_3=0$ and $\beta=1$.
This simplified model has been instrumental in the biological understanding of the Warburg effect \cite{warburg1956origin}, where fast growing cancer cells engage in the apparently wasteful activity of secreting copious amounts of lactate even in the presence of oxygen \cite{vazquez2010catabolic,cossio2017ploscb,cossio2017limits,cossio2018physical,vazquez2017book}. 
We assume that cells are coupled by the byproduct of the third reaction only, so that $J_i=\Delta_i=0$ for $i=1,2$.
In this case, the RS solution simplifies to:
\begin{equation}
\begin{aligned}
f &= - \frac{J}{2} m^2 - \frac{\Delta^2}{4} (\zeta^2 - q^2) \\
&\qquad+ \int_{- \infty}^{\infty} \frac{\mathrm{d} t}{\sqrt{2 \pi}} \mathrm{e}^{- t^2 /2}
\ln \mathrm{Tr} \exp (-\mathcal{H}^{\mathrm{RS}})
\end{aligned}
\end{equation}
where
\begin{equation}
\begin{aligned}
\mathcal{H}^{\mathrm{RS}} &= h (v_1 - v_2) + \left( J m + \Delta \sqrt{q} t \right) (v_1 + v_2) \\
&\qquad- \frac{\Delta^2}{2} (v_1 + v_2)^2 (\zeta - q)
\end{aligned}
\end{equation}
and $m,q,\zeta$ are such that $f$ is extremized.
Upon differentiation the set of coupled equations for the order parameters takes the form:
\begin{equation}
m=\overline{\langle v_3\rangle},\quad
q=\overline{\langle v_3\rangle^2},\quad
\zeta=\overline{\langle v_3^2\rangle}
\label{eq:order-av}
\end{equation}
where the angle brackets denote a Boltzmann average with Hamiltonian $\mathcal H^\mathrm{RS}$ at fixed $t$, and the overbar denotes a Gaussian average over $t$.
Alternatively, \eqref{eq:order-av} can be interpreted as giving the statistics of flux $v_3$ for a given cell, with the angle brackets denoting a Boltzmann average at fixed disorder and the overbar the average over the disorder.

This model exhibits three distinct phases, depending on the interactions between the cells, ($J,\Delta$). Figure \ref{fig:sgphases} shows the types of solutions obtained. When $m>0$ ($m<0$), the population exhibits an overall net flux of production (consumption) of the product of the third reaction, the ferromagnetic state. 
But it is also possible that cells arrange themselves in such a way that the average flux $v_3$ balances, with no bias for production or consumption at the level of the population. In this case $m=0$, and  the parameter $q$ allows us to draw a further distinction. If $q=0$, each cell in the population has an average flux of $v_3=0$, a bona-fide paramagnetic state.
On the other hand, if $q>0$, although the net flux over the entire population is zero, different cells in the population will have positive or negative average values of $v_3$.
This is analogous to the \emph{spin glass} phase of statistical mechanics \cite{mezard1987book}.
Biologically, cells spontaneously specialize into different metabolic phenotypes that cooperate through the exchange of the byproduct.


\begin{figure}
\centerline{\includegraphics[width=0.5\textwidth]{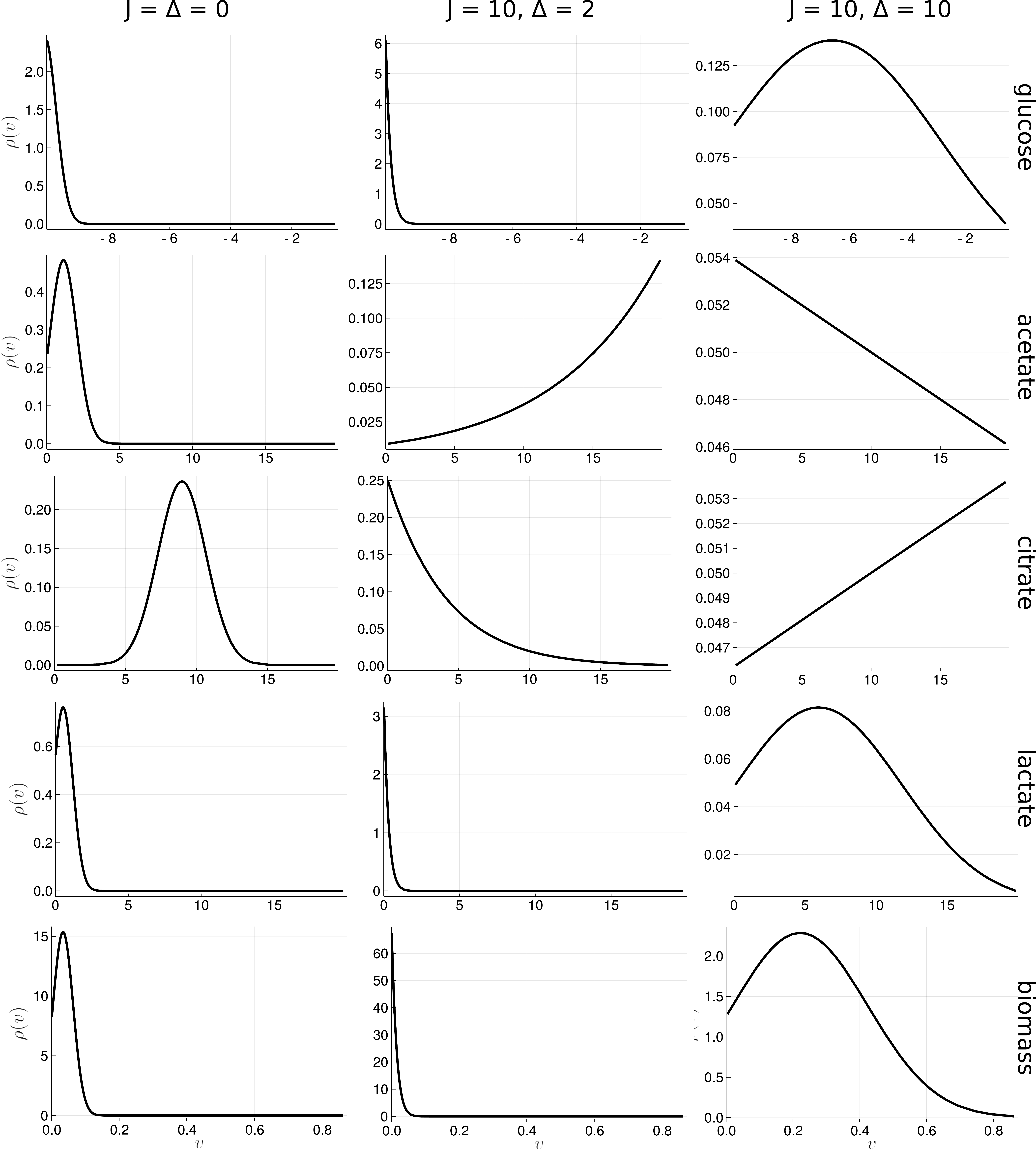}}
\caption{\label{fig:ecoliflux} \emph{E. coli} core metabolic network.
Population flux histograms for selected reactions of the network, for different values of the disorder average ($J$) and fluctuations ($\Delta$). From top to bottom, glucose, acetate, citrate, lactate and biomass.}
\end{figure}

\begin{figure}
\centerline{\includegraphics[width=0.5\textwidth]{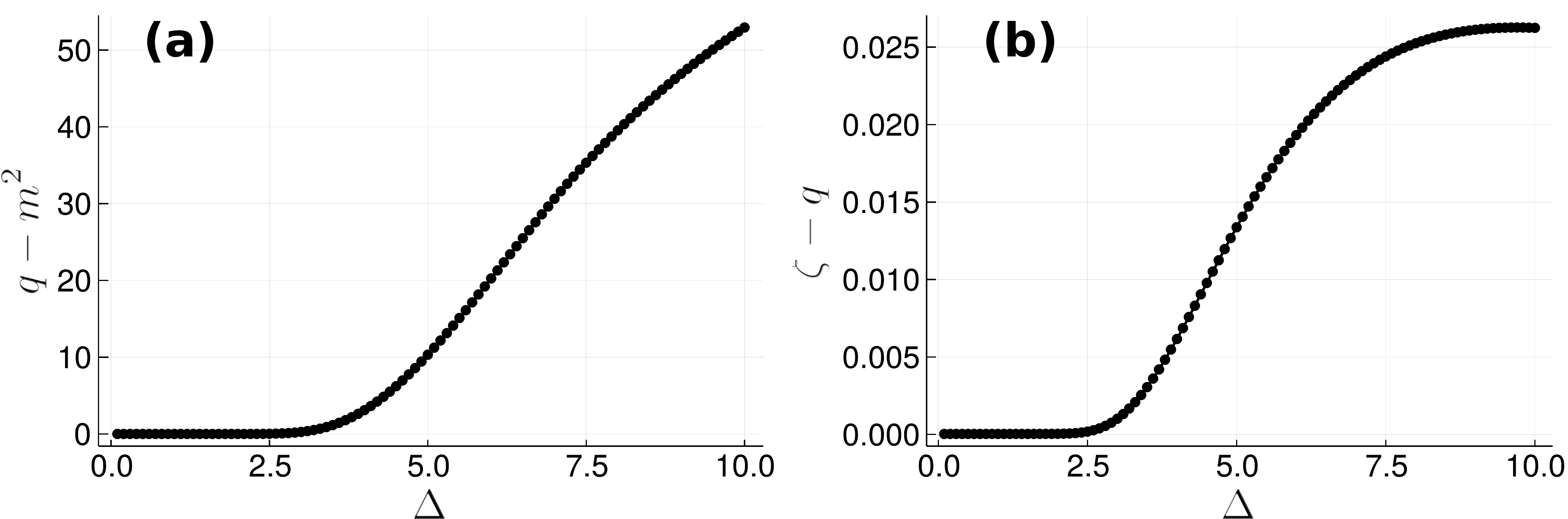}}
\caption{\label{fig:ecoli} \emph{E. coli} core metabolic network.
\textbf{(a)} Expected metabolic variance between samples with different realizations of disorder.
\textbf{(b)} Expected metabolic variance in population with a given realization of the disorder.}
\end{figure}

A more realistic setting, usually exploited  to study the metabolism of bacteria, is the core metabolic network of the \emph{E. coli}. This network consists of 72 metabolites and 87 reactions \cite{ecolicore}.
For simplicity we assume here that the cells are only positively coupled by the exchange of acetate. This may happen in long-term cultures \cite{treves1998acetate} where interactions mediated by acetate, such as cross-feeding polymorphisms appear \cite{treves1998acetate}, or when spatial structure influences the evolution of cooperative and competitive phenotypes \cite{bachmann2013availability}.

For large metabolic networks like this, the computation of the Trace in (\ref{eq:freeenergy}) becomes computationally challenging.
It is a  multivariate integral over a high-dimensional polytope that can be approximately solved using the Expectation Propagation algorithm (see Supplementary Materials) \cite{minka2001expectation,braunstein2017nc,cossio2019maxent}.
We simulated this network with its original flux bounds \cite{ecolicore} and with mean couplings of $J=10$ in the acetate exchange flux and with $h=10$ for the Biomass synthesis reaction, accounting for an evolutionary pressure towards fast growth.
These parameter values are chosen to be on the same order as flux values in the default units of the original network.
Finally we set $\beta=1$.


To illustrate how the metabolic profile  of the population is affected by the interactions in Figure \ref{fig:ecoliflux} we plot  the average distribution function characterizing selected fluxes of the network for different values of the parameters of the model (see Supplementary Materials). As reference, we plot in the first column the distribution of these fluxes in the absence of interactions. In a presence of interaction, for a weakly disordered system,  $\Delta=2$ (second column), cells start to produce more acetate, at the expense of a lower biomass production. In this case, energetic metabolism is down-regulated in the majority of cells. This is exemplified by the reduction in the production of citrate, a reaction of the Krebs cycle, and the fermentation to lactate.

As the disorder increases (third column), $\Delta=10$, interactions mediated by acetate become less dominant in the metabolic fate of the cells, which begin to divert resources towards other pathways at the expense of acetate production. Glucose consumption spreads towards lower velocities, but citrate synthase has a higher flux indicating that energy is being produced by efficient metabolic routes. Then, Lactate production and the growth rate increase, making the culture more biosynthetically active than independent cells in the leftmost column. Therefore, while at low disorder cells focus on the production of acetate, higher disorder shifts the population towards the default metabolic modes of generating energy and biomass synthesis. 

However, a more complex picture emerges looking into the order parameters of the model, defined in the same way as in the simple network solved above.
In Fig. \ref{fig:ecoli}(a) we show the dependence of $q-m^2$ with $\Delta$.
For small $\Delta$, the disorder has no effect on the behavior of the population and the sample to sample fluctuations in the average acetate flux of a given cell vanish. For $\Delta\approx3$, a phase transition occurs and sample to sample fluctuations become important.
On the other hand, Fig. \ref{fig:ecoli}(b) shows the expected metabolic variance in the acetate flux of a cell for a given instance of the disorder.
If $\Delta$ is small, this variance is negligible.
Above the phase transition, the variances become positive, different cells use acetate in different ways and the interpretation of the third column in Fig. \ref{fig:ecoliflux} should be done with care.
This is an average over the disorder and, in this phase, does not represent the typical behavior of a population in a particular sample. It misses the actual heterogeneity of the population.  As in the simplified model, in the statistical mechanics jargon, for $\Delta>3$ the system is in a spin glass phase \cite{mezard1987book}.

\section*{Discussion}

We  provide a general scenario to deal with interacting cells taking into account the real complexity of their metabolism.
Within this scenario, cells may compete for the same specific nutrients, and by-products of one cell can be used by or be toxic to others. This {\it competition} or {\it cooperation} is defined by the interaction between the cells {\em and} by their actual metabolic  capabilities. The later are fixed by the biochemical constraints and imposed by the stochiometric matrix, i.e. the flux conservation within the cell.

We solved the problem within a mean field approximation mimicking a structureless environment.  This solution is summarized in full generality (within the RS formalism as the saddle point of Eq. (3) (see also the Supplementary Materials). 
To first explore its implications we started studying a very simple metabolic network, representing in a highly ideal manner the two major metabolic modes of a mammalian cell: fermentation and respiration. We showed that three qualitatively distinct phases are very well defined. A disordered (paramagnetic phase), in which the interaction between the cells is very small ($J$ and $\Delta \ll 1$) and they behave independently, optimizing their own growth. For very homogeneous interactions $J\gg1$ the cells (that in this case compete for the same nutrient) share the same metabolic state (that can be interpreted as a ferromagnetic state), which can be either one where all the cells in the population ferment, or respire, depending on the initial conditions.
When the interaction has large variations from cell to cell ($\Delta \gg J$), a spin-glass phase appears.
In this toy model the emergence of these phases is uniquely determined by two parameters, $J$ and $\Delta$  but different constraints in this metabolic network would change the form of the phase diagram. This capacity to deal with the specificities of the internal metabolism is what make this approach to the study of interacting cells so appealing.

The solution can be extended to genome scale metabolic networks. Technically the only challenging part is the computation of the Trace over fluxes in Eq. \eqref{eq:freeenergy}.
This could be cumbersome for complex networks, but in the last few years the scientific community have made important progresses in this direction \cite{braunstein2017nc,martino2015mc,cossio2016bp,braunstein2008bmc}. Exploiting the Expectation Propagation algorithm \cite{braunstein2017nc,cossio2019maxent} we were able to solve our model numerically using the \emph{E. coli} core metabolic network. In this more realistic setting, we confirmed qualitatively the results obtained in the simplified model.


\begin{figure}
\centerline{\includegraphics[width=0.5\textwidth]{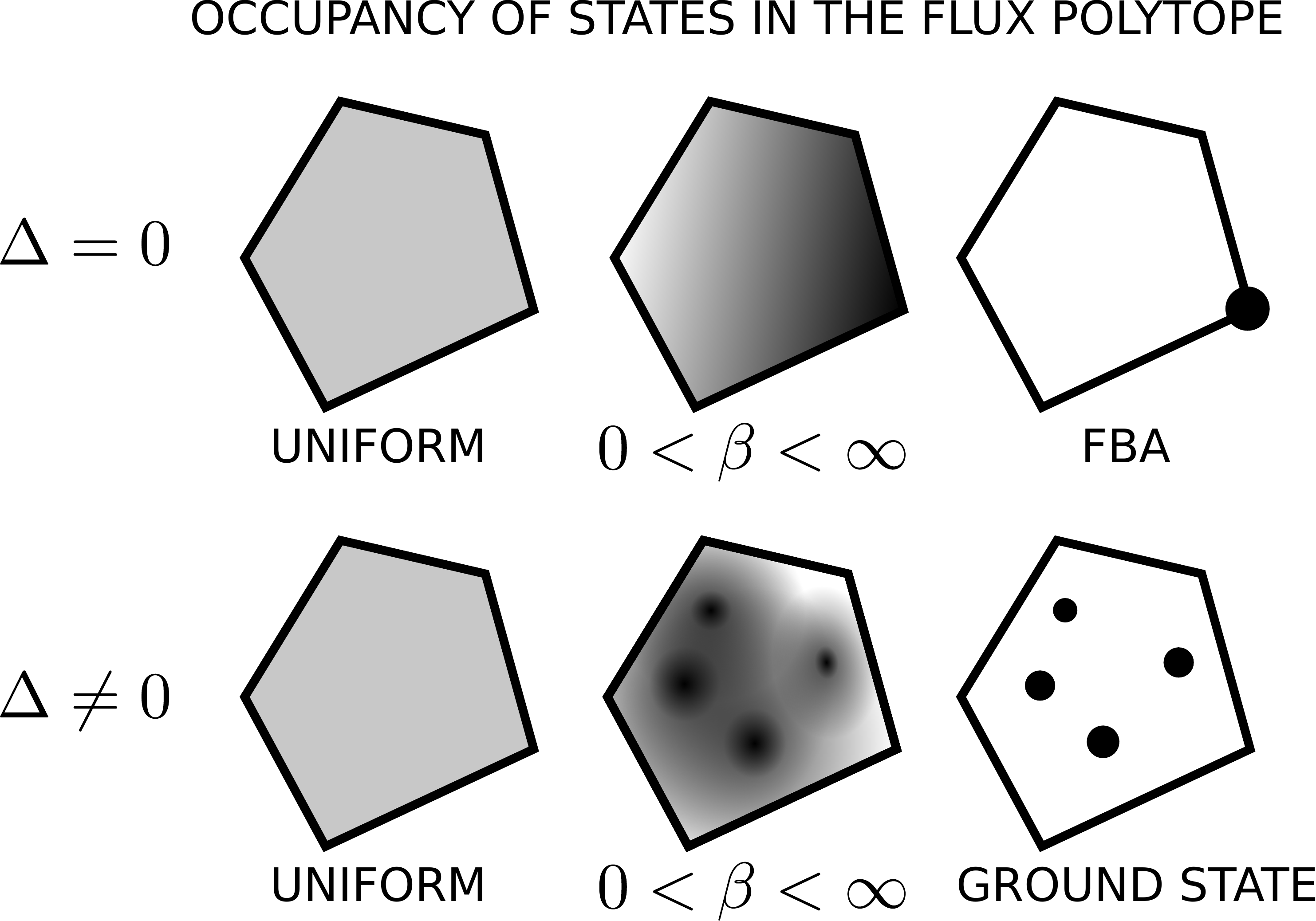}}
\caption{Stoichiometric and reversibility constrains define a high-dimensional polytope of feasible metabolic fluxes for each cell, here represented as a 2-dimensional polygon. In absence of interactions ($\Delta=0$), the distribution of cells within this space has a peak at the maximum growth rate. As $\beta\rightarrow\infty$ the cells concentrate sharply at this peak (FBA), while if $\beta=0$ the cells diffuse over the entire space uniformly (UNIFORM) \cite{martino2018natcomm, cossio2019maxent}. When cells interact ($\Delta\ne0$), there might be multiple peaks which collectively define the ground state of the model. At $\beta\rightarrow\infty$ the cells concentrate sharply at these peaks with different masses. At $\beta=0$ we recover the uniform distribution, while for $0<\beta<\infty$, the cells form diffuse clouds around the peaks. 
\label{fig:polytope}}
\end{figure}

Once the cells are fixed in a tissue, or a culture, mean-field approximations give a general clue about the general physics behind the model of interest, but may fail to catch the whole richness of a problem. 
This, for example, was certainly the case for second order phase transitions \cite{mezard1987book} and we expect that something similar may happen in this scenario. 
For these kinds of problems, statistical physics rests mainly on simulations and we follow  a similar approach here modeling a two dimensional collection of interacting cells.  In the Supplementary Materials we show results for simulations of our toy model in a finite dimensional lattice. Again, we obtain a rich picture, consistent with our mean-field calculations. The metabolism of the cells within the tissue may have a very disordered structure (paramagnetic or spin glass)  or may be organized in an anti-ferromagnet structure, in which by-products of one cell are used by neighboring ones. Such a scenario has been observed in cancer tissues exchanging metabolites with stromal cells \cite{brooks2000intra,cossio2017micro}, and also in healthy tissues such as the brain, where astrocytes echange lactate with neurons \cite{pellerin1998evidence,magistretti2018brain,belanger2011brain}. 

In conclusion, the general picture emerging from our results can be summarized as follows ( see Fig. \ref{fig:polytope}). In absence of interactions or selective pressure, cells distribute uniformly over the space of metabolic states allowed by physico-chemical constrains. The selective pressure tends to favor a concentration of cells near the state of maximum fitness, which in this case corresponds to the FBA solutions \cite{martino2018natcomm,cossio2019maxent}. But when selective pressure is combined with the interaction between the cells, the population may reach equilibrium states where the cells concentrate in different points of the phenotypic space. Different cells may specialize in different functions establishing relations of cooperation or competition between them.

\section*{Acknowledgements}

This project has received funding from the European Union’s Horizon 2020 research and innovation programme MSCA-RISE-2016 under grant agreement No. 734439 INFERNET.

\section*{Author contributions}

All authors contributed equally to this work.

\section*{Competing financial interests}

The authors declare no competing financial interests.

\clearpage

\bibliographystyle{unsrt}
\bibliography{metabsglass}

\end{document}


\title{
  Spin glass theory of metabolic networks
  \thanks{\textit{Note:} {\withTeXmacstext}}
  \thanks{\textit{Note:} {\withTeXmacstext}}
}

\author{
  Jorge Fernandez-de-Cossio-Diaz \& Roberto Mulet
  \\
  \textit{Email:} \texttt{j.cossio.diaz@gmail.com}
}

\date{September 1, 2019}

\maketitle

\section{Model}

We consider a population of $N$ cells, described by the variables $v_{ri}$,
for $r = 1, \ldots, R$ and $i = 1, \ldots, N$, giving the flux of reaction $r$
in cell $i$. These fluxes are subject to $M$ linear constrains (metabolites),
$\sum_r S_{m r} v_{r i} = C_m$ for $m = 1, \ldots, M$, and inequalities
$\mathrm{lb}_r \le v_{r i} \le \mathrm{ub}_r$ for all $r, i$. In our model the
probability that the fluxes of the population are $\{ v_{ri} \}$ is given by a
Boltzmann distribution $\frac{1}{Z}  \mathrm{e}^{- H (\{ v \})}$, where $H (\{
v \}) = - \sum_{r, i} h_r v_{r i} - \sum_{i < j} \sum_r J_{r i j} v_{r i} v_{r
j}$ is the ``energy'' of a flux configuration in the population of cells, and
$Z =\ensuremath{\operatorname{Tr}} \exp (- H (\{ v \}))$ the partition
function. Here $\{ v \}$ refers to all the metabolic fluxes in the entire cell
population, and $\ensuremath{\operatorname{Tr}}$ is a trace over these
variables:
\begin{equation}
  \ensuremath{\operatorname{Tr}} (\cdot) := \int_{\mathrm{lb}}^{\mathrm{ub}}
  (\cdot) \prod_{m i} \delta \left( \sum_r S_{m r} v_{r i} - C_m \right)
  \prod_{r i} \mathrm{d} v_{r i} \label{fluxtrace}
\end{equation}
Generally we will use $\ensuremath{\operatorname{Tr}}$ to denote a trace like
this over all flux variables appearing to its right, always respecting the
relevant stoichiometric constrains. Here we use a Dirac-delta function to
impose these constrains, but they can also be imposed parametrically. The free
energy density is defined by $\mathfrak{f}= - (1 / N) \ln Z$.

\section{Disorder}

We consider $h_r$ fixed and $J_{r i j}$ drawn independently from normal
distributions with means $J_r / N$ and variances $\Delta_r^2 / N$, that we
denote by:
\[ \mathcal{N}_r (J_{r i j}) = \sqrt{\frac{N}{2 \pi}} \frac{1}{\Delta_r} \exp
   \left[ - \frac{N}{2} \left( \frac{J_{r i j} - J_r}{\Delta_r} \right)^2
   \right] \]
The scaling factors $1 / N$ on the mean and variance guarantee that the energy
is extensive. Due to the disorder, $Z$ becomes a random quantity. We are
interested in the average free energy density over the quenched disorder in
the thermodynamic limit ($N \rightarrow \infty$), defined by
$\bar{\mathfrak{f}} = - \lim_{N \rightarrow \infty}  \frac{1}{N} \overline{\ln
Z}$. Here we use an overbar to indicate an average over the values of $J$:
\[ \frac{1}{N} \overline{\ln Z} = \int \left( \prod_{i < j} \prod_r
   \mathcal{N}_r (J_{r i j}) \mathrm{d} J_{r i j} \right) \ln
   \ensuremath{\operatorname{Tr}} \exp (- H (\{ v \})) \]
In what follows, integrations without limits are unbounded.

\section{Replica method}

To compute $\bar{\mathfrak{f}}$ we employ the replica trick, starting from the
formula $\overline{\ln Z} = \lim_{n \rightarrow 0}  \frac{1}{n} \ln
\overline{Z^n}$. Assuming that $n$ is a positive integer,
\begin{equation}
  \overline{Z^n} = \int \mathrm{d} J\mathcal{N} (J) \mathrm{Tr}_{} 
  \mathrm{e}^{- \sum_{\alpha} H (\{ v^{\alpha} \})} = \mathrm{Tr}_{} 
  \overline{\mathrm{e}^{- \sum_{\alpha} H (\{ v^{\alpha} \})}}
  \label{averageZreplica}
\end{equation}
where $\alpha = 1, \ldots, n$ are replica indexes and $\{ v^{\alpha} \}$ is
the set of reaction fluxes of all cells in replica $\alpha$. Note that
$\ensuremath{\operatorname{Tr}}$ here denotes a configurational trace over the
flux variables in all replicas. Thus $Z^n$ is the partition function $n$
replicas of the system, all with the same instance of the disorder. Averaging
over this common instance of disorder gives $\overline{Z^n}$.

Performing the average over $J$ and neglecting sub-dominant terms in the large
$N$ limit (to exponential first order)
\begin{eqnarray*}
  \overline{\mathrm{e}^{- \sum_{\alpha} H (\{ v^{\alpha} \})}} & = & \prod_r
  \exp \left[ \sum_{i, \alpha} h_r v_{ri}^{\alpha} + \sum_{i < j}
  \sum_{\alpha} \frac{J_r}{N} v_{r i}^{\alpha} v_{r j}^{\alpha} + \sum_{i < j}
  \frac{\Delta_r^2}{2 N} \left( \sum_{\alpha} v_{r i}^{\alpha} v_{r
  j}^{\alpha} \right)^2 \right]\\
  & = & \prod_r \exp \left[ \sum_{i, \alpha} h_r v_{r i}^{\alpha} +
  \sum_{\alpha} \frac{J_r}{2 N} \left( \sum_i v_{ri}^{\alpha} \right)^2 +
  \sum_{\alpha \leqslant \beta} \frac{\Delta_r^2}{2 N} \left( \sum_i v_{r
  i}^{\alpha} v_{r i}^{\beta} \right)^2 \left( 1 - \frac{\delta_{\alpha
  \beta}}{2} \right) \right]
\end{eqnarray*}
To get rid of the squares we employ the Hubbard-Stratonovich transform,
$\mathrm{e}^{a x^2 / N} = \sqrt{\frac{N}{2 \pi}} \int \mathrm{e}^{-
\frac{N}{2} u^2 + \sqrt{2 a} x u} \mathrm{d} u$. Introducing variables
$m_r^{\alpha}, q_r^{\alpha \beta}$ $(\alpha \leqslant \beta)$, for each of the
squared terms, and ignoring sub-exponential factors
\begin{eqnarray}
  \overline{\mathrm{e}^{- \sum_{\alpha} H (\{ v^{\alpha} \})}} & = & \prod_r
  \int \left( \prod_{\alpha} \sqrt{\frac{N}{2 \pi}} \mathrm{d} m_r^{\alpha}
  \prod_{\alpha \leqslant \beta} \sqrt{\frac{N}{2 \pi}} \mathrm{d} q_r^{\alpha
  \beta} \right) \exp \left( - \sum_{\alpha} \frac{N}{2} (m_r^{\alpha})^2 -
  \sum_{\alpha \leqslant \beta} \frac{N}{2} (q_r^{\alpha \beta})^2 \right)
  \nonumber\\
  &  & \times \prod_i \exp \left( \sum_{\alpha} h_r v_{r i}^{\alpha} +
  \sum_{\alpha} \sqrt{J_r} m_r^{\alpha} v_{r i}^{\alpha} + \sum_{\alpha
  \leqslant \beta} \frac{\Delta_r q_r^{\alpha \beta}}{\sqrt{1 + \delta_{\alpha
  \beta}}} v_{r i}^{\alpha} v_{r i}^{\beta} \right)  \label{HeffnHubbard}
\end{eqnarray}
Sub-dominant factors to first exponential order can be disregarded, because
they vanish in the limit $N \rightarrow \infty$ in the computation of the free
energy density. After the Hubbard-Stratonovich transform, the cells $i$ become
independent. Therefore:
\begin{eqnarray*}
  \overline{Z^n} & = & \int \prod_r \left( \prod_{\alpha}
  \mathrm{{\textbf{}}d} m_r^{\alpha} \prod_{\alpha \leqslant \beta} \mathrm{d}
  q_r^{\alpha \beta} \right) \exp \left( - \sum_{\alpha} \frac{N}{2}
  (m_r^{\alpha})^2 - \sum_{\alpha \leqslant \beta} \frac{N}{2} (q_r^{\alpha
  \beta})^2 \right)\\
  &  & \times \exp \left\{ N \ln \ensuremath{\operatorname{Tr}} \prod_r \exp
  \left[ \sum_{\alpha} \left( h_r + \sqrt{J_r} m_r^{\alpha} \right)
  v_r^{\alpha} + \sum_{\alpha \leqslant \beta} \frac{\Delta_r q_r^{\alpha
  \beta}}{\sqrt{1 + \delta_{\alpha \beta}}} v_r^{\alpha} v_r^{\beta} \right]
  \right\}
\end{eqnarray*}
where $\mathrm{Tr}$ performs a trace over the replica single-site variables
$v_r^{\alpha}$. In the $N \rightarrow \infty$ limit we can perform the
integration over the variables $m_{\alpha r}, q_{\alpha \beta r}$ by the
saddle-point method. It follows that, to leading exponential order in $N$:
\[ \frac{1}{N} \ln \overline{Z^n} = - \sum_{r, \alpha}
   \frac{(m_r^{\alpha})^2}{2} - \sum_r \sum_{\alpha \leqslant \beta}
   \frac{(q_r^{\alpha \beta})^2}{2} + \ln \ensuremath{\operatorname{Tr}}
   \prod_r \exp \left[ \sum_{\alpha} \left( h_r + \sqrt{J_r} m_r^{\alpha}
   \right) v_r^{\alpha} + \sum_{\alpha \leqslant \beta} \frac{\Delta_r
   q_r^{\alpha \beta}}{\sqrt{1 + \delta_{\alpha \beta}}} v_r^{\alpha}
   v_r^{\beta} \right] \label{averageZn} \]
where now $m_{\alpha r}, q_{\alpha \beta r}$ are such that the exponent is
extremized. Differentiating we find the saddle-point equations:
\begin{equation}
  m_r^{\alpha} = \frac{\ensuremath{\operatorname{Tr}}
  \mathrm{e}^{-\mathcal{H}} v_r^{\alpha}}{\ensuremath{\operatorname{Tr}}
  \mathrm{e}^{-\mathcal{H}}}, \hspace{3em} q_r^{\alpha \beta} =
  \frac{\ensuremath{\operatorname{Tr}} \mathrm{e}^{-\mathcal{H}} v_r^{\alpha}
  v_r^{\beta}}{\ensuremath{\operatorname{Tr}} \mathrm{e}^{-\mathcal{H}}}
  \qquad (\alpha \leqslant \beta) \label{saddle}
\end{equation}
where \ $\mathcal{H}= - \sum_{r \alpha} (h_r + J_r m_r^{\alpha}) v_r^{\alpha}
- \sum_r \sum_{\alpha \leqslant \beta} \frac{\Delta_r^2 q_r^{\alpha \beta}}{1
+ \delta_{\alpha \beta}} v_r^{\alpha} v_r^{\beta}$, and we made the changes of
variables $m_r^{\alpha} \rightarrow m_r^{\alpha} / \sqrt{J_r}$, $q_r^{\alpha
\beta} \rightarrow q_r^{\alpha \beta} \sqrt{1 + \delta_{\alpha \beta}} /
\Delta_r$.

\section{Replica-symmetric {\itshape{ansatz}}}

We assume that the global maximizer of {\eqref{averageZn}} respects replica
symmetry (RS), {{\em i.e.\/}}, that the maximum occurs at a RS point where
$m_{\alpha r} = m_r$, $q_r^{\alpha \alpha} = \zeta_r$, $q_r^{\alpha \beta} =
q_r$ ($\alpha < \beta$) are the same for all replicas. Substituting in
{\eqref{averageZn}} (after the change of variables $m_r \rightarrow m_r /
\sqrt{J_r}$, $q_r \rightarrow q_r / \Delta_r$, $\zeta_r \rightarrow \zeta_r
\sqrt{2} / \Delta_r$), and dropping vanishing terms in the limit $n
\rightarrow 0$:
\[ \frac{1}{N n} \ln \overline{Z^n} = - \sum_r \left[ \frac{J_r}{2} m_r^2 +_r
   \frac{\Delta_r^2}{4} (\zeta_r^2 - q_r^2) \right] + \frac{1}{n} \ln
   \ensuremath{\operatorname{Tr}} \prod_r \exp \left[ \sum_{\alpha} (h_r + J_r
   m_r) v_r^{\alpha} + \Delta_r^2 q_r \sum_{\alpha < \beta} v_r^{\alpha}
   v_r^{\beta} + \frac{\Delta_r^2 \zeta_r}{2} \sum_{\alpha} (v_r^{\alpha})^2
   \right] \label{averageZnRS} \]
Using the identity $\sum_{\alpha < \beta} v_r^{\alpha} v_r^{\beta} =
\frac{1}{2} \left( \sum_{\alpha} v_r^{\alpha} \right)^2 - \frac{1}{2}
\sum_{\alpha} (v_r^{\alpha})^2$, we obtain $q_r \sum_{\alpha < \beta}
v_r^{\alpha} v_r^{\beta} + \frac{\zeta_r}{2} \sum_{\alpha} (v_r^{\alpha})^2 =
\frac{\zeta_r - q_r}{2} \sum_{\alpha} (v_r^{\alpha})^2 + \frac{q_r}{2} \left(
\sum_{\alpha} v_r^{\alpha} \right)^2$. Employing the Hubbard-Stratonovich
transformation once more, we introduce variables $t_r$ to eliminate the
squared terms $\left( \sum_{\alpha} \phi_r^{\alpha} \right)^2$ from the
exponent:
\begin{eqnarray*}
  \frac{1}{N n} \ln \overline{Z^n} & = & - \sum_r \left[ \frac{J_r}{2} m_r^2
  +_r \frac{\Delta_r^2}{4} (\zeta_r^2 - q_r^2) \right] + \frac{1}{n} \ln
  \ensuremath{\operatorname{Tr}} \prod_r \int \exp \left[ \sum_{\alpha} \left(
  h_r + J_r m_r + \Delta_r \sqrt{q_r} t_r \right) v_r^{\alpha} +
  \frac{\Delta_r^2}{2} (\zeta_r - q_r) \sum_{\alpha} (v_r^{\alpha})^2 \right]
  \mathrm{D} t_r\\
  & = & - \sum_r \left[ \frac{J_r}{2} m_r^2 +_r \frac{\Delta_r^2}{4}
  (\zeta_r^2 - q_r^2) \right] + \frac{1}{n} \ln \int \left\{
  \ensuremath{\operatorname{Tr}} \prod_r \mathrm{e}^{\left( h_r + J_r m_r +
  \Delta_r \sqrt{q_r} t_r \right) v_r^{\alpha} + \frac{\Delta_r^2}{2} (\zeta_r
  - q_r) (v_r^{\alpha})^2} \right\}^n \mathrm{D} \vec{t}
\end{eqnarray*}
where $\mathrm{D} t_r = \frac{1}{\sqrt{2 \pi}} \mathrm{e}^{- t_r^2 / 2}
\mathrm{d} t_r$ is the standard Gaussian measure and $\mathrm{D} \vec{t} =
\prod_r \mathrm{D} t_r$. As $n \rightarrow 0$, we can use the expansions $x^n
\approx 1 + n \ln x$ and $\ln (1 + n x) \approx n x$, to obtain.
\begin{eqnarray*}
  \frac{1}{N n} \ln \overline{Z^n} & \sim & - \sum_r \left[ \frac{J_r}{2}
  m_r^2 +_r \frac{\Delta_r^2}{4} (\zeta_r^2 - q_r^2) \right] + \int \ln
  \left\{ \ensuremath{\operatorname{Tr}} \prod_r \mathrm{e}^{\left( h_r + J_r
  m_r + \Delta_r \sqrt{q_r} t_r \right) v_r + \frac{\Delta_r^2}{2} (\zeta_r -
  q_r) (v_r)^2} \right\} \mathrm{D} \vec{t} \qquad (n \rightarrow 0)
\end{eqnarray*}
At this point the trace over fluxes is only over variables $\{ v_r \}$
corresponding to a single cell, since the replica index has dropped out.
Finally, taking the limit $n \rightarrow 0$:
\[ \lim_{N \rightarrow \infty} \frac{1}{N} \overline{\ln Z} = - \sum_r \left[
   \frac{J_r}{2} m_r^2 + \frac{\Delta_r^2}{4} (\zeta_r^2 - q_r^2) \right] +
   \int \ln (\ensuremath{\operatorname{Tr}}
   \mathrm{e}^{-\mathcal{H}_{\ensuremath{\operatorname{RS}}}}) \mathrm{D}
   \vec{t} \]
where $\mathcal{H}_{\ensuremath{\operatorname{RS}}} = - \sum_r \left( h_r +
J_r m_r + \Delta_r \sqrt{q_r} t_r \right) v_r - \sum_r \frac{\Delta_r^2}{2}
(\zeta_r - q_r) v_r^2$. Recall that $m_r, q_r, \zeta_r$ are such that this
expression is extremized. Differentiating, we obtain the saddle-point
equations:
\begin{equation}
  m_r = \int  \frac{\ensuremath{\operatorname{Tr}}
  \mathrm{e}^{-\mathcal{H}_{\ensuremath{\operatorname{RS}}}}
  v_r}{\ensuremath{\operatorname{Tr}}
  \mathrm{e}^{-\mathcal{H}_{\ensuremath{\operatorname{RS}}}}} \mathrm{D}
  \vec{t}, \quad q_r = \int  \left( \frac{\ensuremath{\operatorname{Tr}}
  \mathrm{e}^{-\mathcal{H}_{\ensuremath{\operatorname{RS}}}}
  v_r}{\ensuremath{\operatorname{Tr}}
  \mathrm{e}^{-\mathcal{H}_{\ensuremath{\operatorname{RS}}}}} \right)^2
  \mathrm{D} \vec{t}, \quad \zeta_r = \int 
  \frac{\ensuremath{\operatorname{Tr}}
  \mathrm{e}^{-\mathcal{H}_{\ensuremath{\operatorname{RS}}}}
  v_r^2}{\ensuremath{\operatorname{Tr}}
  \mathrm{e}^{-\mathcal{H}_{\ensuremath{\operatorname{RS}}}}} \mathrm{D}
  \vec{t} \label{saddleRS}
\end{equation}
See the Appendix for the steps leading to the saddle-point equation for $q_r$.

\section{Flux histograms}

Now we would like to be able to compute more general statistics of the fluxes
in the population of cells. Let $Z_{\phi} = \frac{1}{N} \sum_i
\ensuremath{\operatorname{Tr}} \mathrm{e}^{- H (\{ v \})} \phi (\vec{v}_i)$,
where $\vec{v}_i$ denotes the vector of fluxes of cell $i$ and $\phi$ is an
arbitrary function. We are interested in the ratio $Z_{\phi} / Z =
\left\langle \frac{1}{N} \sum_i \phi (\vec{v}_i) \right\rangle$, where
$\langle \cdot \rangle$ denotes a Botlzmann average, as before. Computing this
ratio will allow us to compute many statistics of the flux distribution in the
population of cells. For example, if we take $\phi (\vec{v}_i) = \delta (v_{i
1} - v)$, where $\delta (\cdot)$ is the Dirac delta function, then $Z_{\phi} /
Z$ is the average fraction of cells that carry flux $v$ in reaction 1.

In the presence of disorder, $Z_{\phi}$ and $Z$ are random quantities. But
their ratio $Z_{\phi} / Z$ is self-averaging in the thermodynamic limit. As
before we denote an average over the disorder by an overbar. To compute
$\overline{Z_{\phi} / Z}$ we employ a variant of the replica trick,
$\overline{Z_{\phi} / Z} = \lim_{n \rightarrow 0} \overline{Z_{\phi} Z^{n -
1}}$, and proceed to compute $\overline{Z_{\phi} Z^{n - 1}}$ as if $n - 1$ was
a non-negative integer. We have:
\begin{eqnarray*}
  \overline{Z_{\phi} Z^{n - 1}} & = & \frac{1}{N} \sum_i \overline{\{
  \ensuremath{\operatorname{Tr}} \mathrm{e}^{- H (\{ v \})} \phi (\vec{v}_i)
  \} \left\{ \ensuremath{\operatorname{Tr}} \mathrm{e}^{- \sum_{\alpha = 1}^{n
  - 1} H (\{ v^{\alpha} \})} \right\}} =\ensuremath{\operatorname{Tr}}
  \overline{\mathrm{e}^{- \sum_{\alpha} H (\{ v^{\alpha} \})}} \phi
  (\vec{v}_1^n)
\end{eqnarray*}
Since all cells become equivalent after averaging over the disorder, we can
replace the population average $\frac{1}{N} \sum_i \phi (\vec{v}_i)$ with just
$\phi (\vec{v}_1)$. Here a sum like $\sum_{\alpha}$ without explicit limits is
assumed to go through the range $1 \leqslant \alpha \leqslant n$. Thus in the
last step we simply renamed the variables $v_{r i}$ of the first trace to
$v_{r i}^n$, which allowed us to merge both traces into a single trace over
all fluxes $v_{r i}^{\alpha}$ in all replicas, $\alpha = 1, \ldots, n$.

Using {\eqref{HeffnHubbard}},
\begin{eqnarray*}
  \overline{Z_{\phi} Z^{n - 1}} & = & \int \prod_r \left( \prod_{\alpha}
  \mathrm{d} m_r^{\alpha} \prod_{\alpha \leqslant \beta} \mathrm{d}
  q_r^{\alpha \beta} \right) \exp \left( - \sum_{\alpha} \frac{N}{2}
  (m_r^{\alpha})^2 - \sum_{\alpha \leqslant \beta} \frac{N}{2} (q_r^{\alpha
  \beta})^2 \right)\\
  &  & \times \ensuremath{\operatorname{Tr}} \phi (\vec{v}_1^n) \prod_{r i}
  \exp \left( \sum_{\alpha} \left( h_r + \sqrt{J_r} m_r^{\alpha} \right) v_{r
  i}^{\alpha} + \sum_{\alpha \leqslant \beta} \frac{\Delta_r q_r^{\alpha
  \beta}}{\sqrt{1 + \delta_{\alpha \beta}}} v_{r i}^{\alpha} v_{r i}^{\beta}
  \right)\\
  & = & \int \prod_r \left( \prod_{\alpha} \mathrm{d} m_r^{\alpha}
  \prod_{\alpha \leqslant \beta} \mathrm{d} q_r^{\alpha \beta} \right) \exp
  \left( - \sum_{\alpha} \frac{N}{2} (m_r^{\alpha})^2 - \sum_{\alpha \leqslant
  \beta} \frac{N}{2} (q_r^{\alpha \beta})^2 \right)\\
  &  & \times \exp \left\{ N \ln \ensuremath{\operatorname{Tr}} \prod_r \exp
  \left( \sum_{\alpha} \left( h_r + \sqrt{J_r} m_r^{\alpha} \right)
  v_r^{\alpha} + \sum_{\alpha \leqslant \beta} \frac{\Delta_r q_r^{\alpha
  \beta}}{\sqrt{1 + \delta_{\alpha \beta}}} v_r^{\alpha} v_r^{\beta} \right)
  \right\}\\
  &  & \times \frac{\ensuremath{\operatorname{Tr}} \phi (\vec{v}^n) \prod_r
  \exp \left( \sum_{\alpha} \left( h_r + \sqrt{J_r} m_r^{\alpha} \right)
  v_r^{\alpha} + \sum_{\alpha \leqslant \beta} \frac{\Delta_r q_r^{\alpha
  \beta}}{\sqrt{1 + \delta_{\alpha \beta}}} v_r^{\alpha} v_r^{\beta}
  \right)}{\ensuremath{\operatorname{Tr}} \prod_r \exp \left( \sum_{\alpha}
  \left( h_r + \sqrt{J_r} m_r^{\alpha} \right) v_r^{\alpha} + \sum_{\alpha
  \leqslant \beta} \frac{\Delta_r q_r^{\alpha \beta}}{\sqrt{1 + \delta_{\alpha
  \beta}}} v_r^{\alpha} v_r^{\beta} \right)}
\end{eqnarray*}
This integration is dominated by the same saddle-point as in the computation
of $\overline{Z^n}$ above. Therefore, making the change of variables
$m_r^{\alpha} \rightarrow m_r^{\alpha} / \sqrt{J_r}$, $q_r^{\alpha \beta}
\rightarrow q_r^{\alpha \beta} \sqrt{1 + \delta_{\alpha \beta}} / \Delta_r$,
\[ \overline{Z_{\phi} Z^{n - 1}} \propto \ensuremath{\operatorname{Tr}} \phi
   (\vec{v}^n) \prod_r \exp \left( \sum_{\alpha} (h_r + J_r m_r^{\alpha})
   v_r^{\alpha} + \sum_{\alpha \leqslant \beta} \frac{\Delta_r^2 q_r^{\alpha
   \beta}}{1 + \delta_{\alpha \beta}} v_r^{\alpha} v_r^{\beta} \right) \]
where we ignore constant factors independent of $\phi$, and $m_r^{\alpha},
q_r^{\alpha \beta}$ are the solution of the saddle-point equations
{\eqref{saddle}} above. The constant factor can be determined by
normalization.

As before, we consider a replica symmetric {\itshape{ansatz}} for the solution
of the saddle-point equations. Denoting $q_r = q_r^{\alpha \beta}$ ($\alpha <
\beta$) and $\zeta_r = q_r^{\alpha \alpha}$, it follows that:
\begin{eqnarray*}
  \overline{Z_{\phi} Z^{n - 1}} & \propto & \ensuremath{\operatorname{Tr}}
  \phi (\vec{v}^n) \prod_r \exp \left( (h_r + J_r m_r) \sum_{\alpha}
  v_r^{\alpha} + \Delta_r^2 q_r \sum_{\alpha < \beta} v_r^{\alpha} v_r^{\beta}
  + \frac{\Delta_r^2 \zeta_r}{2} \sum_{\alpha} (v_r^{\alpha})^2 \right)\\
  & = & \ensuremath{\operatorname{Tr}} \phi (\vec{v}^n) \prod_r \exp \left(
  (h_r + J_r m_r) \sum_{\alpha} v_r^{\alpha} + \frac{\Delta_r^2}{2} q_r \left(
  \sum_{\alpha} v_r^{\alpha} \right)^2 + \frac{\Delta_r^2}{2} (\zeta_r - q_r)
  \sum_{\alpha} (v_r^{\alpha})^2 \right)
\end{eqnarray*}
Employing the Hubbard-Stratonovich transformation, we introduce Gaussian
variables $t_r$ to eliminate the square:
\begin{eqnarray*}
  \overline{Z_{\phi} Z^{n - 1}} & \propto & \ensuremath{\operatorname{Tr}}
  \phi (\vec{v}^n) \int \mathrm{D} \vec{t}  \prod_{\alpha, r} \exp \left(
  \left( h_r + J_r m_r + \Delta_r \sqrt{q_r} t_r \right) v_r^{\alpha} +
  \frac{\Delta_r^2}{2} (\zeta_r - q_r) (v_r^{\alpha})^2 \right)\\
  & = & \int \frac{\ensuremath{\operatorname{Tr}} \phi (\vec{v})
  \mathrm{e}^{-\mathcal{H}_{\ensuremath{\operatorname{RS}}}}}{[\ensuremath{\operatorname{Tr}}
  \mathrm{e}^{-\mathcal{H}_{\ensuremath{\operatorname{RS}}}}]^{1 - n}}
  \mathrm{D} \vec{t}
\end{eqnarray*}
where $\mathcal{H}_{\ensuremath{\operatorname{RS}}} = - \sum_r \left( h_r +
J_r m_r + \Delta_r \sqrt{q_r} t_r \right) v_r - \sum_r \frac{\Delta_r^2}{2}
(\zeta_r - q_r) v_r^2$, as before. Finally, taking the limit $n \rightarrow
0$:
\begin{eqnarray}
  \overline{Z_{\phi} / Z} & = & \int \frac{\ensuremath{\operatorname{Tr}} \phi
  (\vec{v})
  \mathrm{e}^{-\mathcal{H}_{\ensuremath{\operatorname{RS}}}}}{\ensuremath{\operatorname{Tr}}
  \mathrm{e}^{-\mathcal{H}_{\ensuremath{\operatorname{RS}}}}} \mathrm{D}
  \vec{t}  \label{Z/Zave}
\end{eqnarray}
The underdetermined constant from the previous steps does not depend on
$\phi$. Since $Z_{\phi} / Z = 1$ when $\phi (\vec{v}) = 1$, the value of this
constant must be 1, which is why {\eqref{Z/Zave}} becomes an equation and not
just a proportionality relation.

Now we take $\phi (\vec{v}) = \delta (v_r - v)$ and denote $\overline{\rho_r
(v)} := \overline{Z_{\phi} / Z}$ for this choice of $\phi$. As mentioned
before, $\rho_r (v)$ is the average fraction of cells with the $r$'th equal to
$v$. In this case, {\eqref{Z/Zave}} reduces to
\begin{eqnarray}
  \overline{\rho_r (v)} & = & \int \frac{\ensuremath{\operatorname{Tr}} \delta
  (v_r - v)
  \mathrm{e}^{-\mathcal{H}_{\ensuremath{\operatorname{RS}}}}}{\ensuremath{\operatorname{Tr}}
  \mathrm{e}^{-\mathcal{H}_{\ensuremath{\operatorname{RS}}}}} \mathrm{D}
  \vec{t}  \label{rhov}
\end{eqnarray}
With this expression we can compute the marginal distribution of a reaction
flux in the population of cells with random disorder.

\section{Expectation propagation}

The trace over fluxes $\ensuremath{\operatorname{Tr}} (\cdot)$ was defined in
{\eqref{fluxtrace}} as an integration over the polytope of fluxes, subject to
stoichiometric constrains and flux bounds. For large metabolic networks this
integration is intractable, but approximations are possible. We exploit here
the expectation propagation (EP) algorithm of {\cite{braunstein2017ep}}. We
start with replacing the hard stoichiometric constrains implicit in the trace
with softer Gaussian constrains. Define the probability distribution:
\[ P_{\beta} (\vec{v}) \propto \mathrm{e}^{- \frac{\beta}{2} \| \mathbb{S}
   \vec{v} - \vec{C} \|^2} \prod_r \psi_r (v_r), \qquad
   \ensuremath{\operatorname{Tr}}_{\beta} (\cdot) = \int (\cdot) P_{\beta}
   (\underline{v}) \mathrm{d} \underline{v} \]
where $\psi_r (v_r) = 1$ if $v_r^{\ensuremath{\operatorname{lb}}} \le v_r \le
v_r^{\ensuremath{\operatorname{ub}}}$ and $\psi_r (v_r) = 0$ otherwise.
Eventually we take the limit $\beta \rightarrow \infty$ (in practice we use a
very large value of $\beta$), which reproduces the original definition of the
trace with its hard constrains. Since $P_{\beta} (\vec{v})$ is normalized to
1, note that $\ensuremath{\operatorname{Tr}}_{\beta} (\cdot)$ and
$\ensuremath{\operatorname{Tr}} (\cdot)$ differ in a constant (the volume of
the poplytope), but this is not important because this constant cancels in all
the expressions of interest below.

Dealing with $P_{\beta} (\vec{v})$ directly is still hard due to the
multi-dimensional truncation. Therefore, we further approximate $P_{\beta}
(\vec{v})$ with a multivariate Gaussian $Q (\vec{v})$ of the form:
\[ Q (\vec{v}) \propto \mathrm{e}^{- \frac{\beta}{2} \| \mathbb{S} \vec{v} -
   \vec{C} \|^2} \prod_r \phi_r (v_r), \qquad \phi_r (v_r) \propto \exp \left[
   - \frac{(v_r - a_r)^2}{2 d_r} \right] \]
replacing the flux truncations $\psi_r (v_r)$ with soft univariate Gaussians
$\phi_r (v_r)$. A good set of values for the parameters $\{ a_r, d_r \}$ can
be obtained by the EP algorithm, that we describe briefly. Let $\mathbb{D}$
denote the diagonal matrix with entries $D_{r r} = d_r^{- 1}$. Then $Q
(\underline{v})$ can also be written in the form:
\[ Q (\underline{v}) \propto \exp \left( - \frac{1}{2} (\vec{v} -
   \vec{\mu}_Q)^{\top} \Sigma_Q^{- 1} (\vec{v} - \vec{\mu}_Q) \right)^{},
   \qquad \Sigma_Q^{- 1} = \beta \mathbb{S}^{\top} \mathbb{S}+\mathbb{D},
   \qquad \vec{\mu}_Q = \Sigma_Q (\beta \mathbb{S}^{\top} \vec{C} +\mathbb{D}
   \vec{a}) \]
From this it is easy to obtain the marginal $Q (v_r) = \int Q (\vec{v})
\mathrm{d} \vec{v}_{\backslash r}$, which is a univariate Gaussian with mean
$(\mu_Q)_r$ and variance $(\Sigma_Q)_{r r}$. In EP we approximate the exact
marginal $P_{\beta} (v_r) = \int P_{\beta} (\vec{v}) \mathrm{d}
\vec{v}_{\backslash r}$ by the truncated univariate Gaussian $Q^{(r)} (v_r)$:
\[ Q^{(r)} (v_r) \propto \frac{\psi_r (v_r)}{\phi_r (v_r)} Q (v_r) \propto
   \psi_r (v_r) \exp \left[ - \frac{(v_r - \mu_r)^2}{2 \nu_r} \right], \qquad
   \frac{1}{\nu_r} = \frac{1}{(\Sigma_Q)_{r r}} - \frac{1}{d_r}, \quad
   \frac{\mu_r}{\nu_r} = \frac{(\mu_Q)_r}{(\Sigma_Q)_{r r}} - \frac{a_r}{d_r}
\]
The EP algorithm consists of matching the expected values of $v_r$ and $v_r^2$
under $Q (v_r)$ and $Q^{(r)} (v_r)$, solving the following equations: $\langle
v_r \rangle_{Q^{(r)}} = \langle v_r \rangle_Q$ and $\langle v_r^2
\rangle_{Q^{(r)}} = \langle v_r^2 \rangle_Q$. This gives a number of equations
that is enough to determine the parameters $\{ a_r, d_r \}$. Having determined
$Q (\vec{v})$, the trace over fluxes in the polytope reduces under the EP
approximation to a simple multivariate Gaussian integration in $\vec{v}$, with
truncation in at most one dimension.

Next, $Q_{\ensuremath{\operatorname{RS}}} (\vec{v} | \vec{t}) \propto Q
(\vec{v}) \mathrm{e}^{-\mathcal{H}_{\ensuremath{\operatorname{RS}}}}$ is also
a multivariate Gaussian in $\vec{v}$, for fixed $\vec{t}$ and fixed values of
the order parameters $m_r, q_r, \zeta_r$:
\[ Q_{\ensuremath{\operatorname{RS}}} (\vec{v} | \vec{t}) \propto \exp \left[
   - \frac{1}{2} (\vec{v} - \vec{\mu}_{\ensuremath{\operatorname{RS}}})^{\top}
   \Sigma_{\ensuremath{\operatorname{RS}}}^{- 1} (\vec{v} -
   \vec{\mu}_{\ensuremath{\operatorname{RS}}}) \right], \qquad
   \Sigma_{\ensuremath{\operatorname{RS}}}^{- 1} = \Sigma_Q^{- 1} -\mathbb{G},
   \qquad \vec{\mu}_{\ensuremath{\operatorname{RS}}} =
   \Sigma_{\ensuremath{\operatorname{RS}}} (\Sigma_Q^{- 1} \vec{\mu}_Q +
   \vec{h} + \vec{\chi}) \]
where $\mathbb{G}$ is a diagonal matrix with entries $G_{r r} = \Delta_r^2
(\zeta_r - q_r)$ and $\chi_r = J_r m_r + \Delta_r \sqrt{q_r} t_r$. The
marginal $Q_{\ensuremath{\operatorname{RS}}} (v_r | \vec{t})$ is a univariate
Gaussian with mean $(\mu_{\ensuremath{\operatorname{RS}}})_r$ and variance
$(\Sigma_{\ensuremath{\operatorname{RS}}})_{r r}$. As before, we define the
truncated univariate Gaussians $Q_{\ensuremath{\operatorname{RS}}}^{(r)} (v_r
| \vec{t}) \propto Q_{\ensuremath{\operatorname{RS}}} (v_r | \vec{t})
\frac{\psi_r (v_r)}{\phi_r (v_r)}$ to approximate the flux marginals,
\[ Q_{\ensuremath{\operatorname{RS}}}^{(r)} (v_r | \underline{t}) \propto
   \psi_r (v_r) \exp \left[ - \frac{(v_r - \alpha_r)^2}{2 \tau_r} \right],
   \qquad \frac{1}{\tau_r} =
   \frac{1}{(\Sigma_{\ensuremath{\operatorname{RS}}})_{r r}} - \frac{1}{d_r},
   \qquad \frac{\alpha_r}{\tau_r} =
   \frac{(\mu_{\ensuremath{\operatorname{RS}}})_r}{(\Sigma_{\ensuremath{\operatorname{RS}}})_{r
   r}} - \frac{a_r}{d_r} \]
Note that $\tau_r$ can become negative, in which case
$Q_{\ensuremath{\operatorname{RS}}}^{(r)} (v_r | \underline{t})$ is not a
typical Gaussian. When $\tau_r$ goes from positive to negative values,
$Q_{\ensuremath{\operatorname{RS}}}^{(r)} (v_r | \underline{t})$ starts to
favor flux values far from $\alpha_r$. In this case a finite truncation range
is essential because otherwise $Q_{\ensuremath{\operatorname{RS}}}^{(r)} (v_r
| \underline{t})$ cannot be normalized.

Now we can approximately replace the intractable exact trace ratios of the
form $\frac{\ensuremath{\operatorname{Tr}} (\cdot)
\mathrm{e}^{-\mathcal{H}_{\ensuremath{\operatorname{RS}}}}}{\ensuremath{\operatorname{Tr}}
\mathrm{e}^{-\mathcal{H}_{\ensuremath{\operatorname{RS}}}}}$ in the previous
sections with expectations over $Q_{\ensuremath{\operatorname{RS}}}$ (or the
truncated marginals $Q_{\ensuremath{\operatorname{RS}}}^{(r)}$) and the
Gaussian auxiliary variables $\vec{t}$. The saddle-point equations for the
order parameters are approximated by:
\[ m_r \approx \int v_r Q_{\ensuremath{\operatorname{RS}}}^{(r)} (v_r |
   \vec{t}) \mathrm{d} v_r \mathrm{D} \vec{t}, \quad q_r \approx \int \left(
   \int v_r Q_{\ensuremath{\operatorname{RS}}}^{(r)} (v_r | \vec{t})
   \mathrm{d} v_r \right)^2 \mathrm{D} \vec{t}, \quad \zeta_r \approx \int
   v_r^2 Q_{\ensuremath{\operatorname{RS}}}^{(r)} (v_r | \vec{t}) \mathrm{d}
   v_r \mathrm{D} \vec{t} \]
Having determined the order parameters by fixed-point iteration of these
equations, the marginal flux histograms {\eqref{rhov}} can be approximated
under EP by $\rho_r (v_r) \approx \int
Q_{\ensuremath{\operatorname{RS}}}^{(r)} (v_r | \vec{t}) \mathrm{D} \vec{t}$,
as shown above.

\section{Single reaction coupling}

As an example, let us see how this framework simplifies in the case where
cells are coupled by a single reaction. Without any loss of generality we can
assume here that this coupled reaction has index $r = 1$, so that $\Delta_r =
J_r = 0$ for all $r > 1$ but $\Delta_1, J_1$ can be non-zero. To lighten the
notation we drop the index, and denote $m_1 = m$, $q_1 = q$, $\zeta_1 =
\zeta$, $t_1 = t$, $\Delta_1 = \Delta$, $J_1 = J$. Then $\mathbb{G}= \Delta^2
(\zeta - q) \vec{e}_1 \vec{e}_1^{\top}$ and $\vec{\chi} = \left( J m + \Delta
\sqrt{q} t \right) \vec{e}_1$, where $\vec{e}_1$ is the unit vector in the
direction of the coupled reaction. Since
$\Sigma_{\ensuremath{\operatorname{RS}}}^{- 1} = \Sigma_Q^{- 1} -\mathbb{G}$
is a rank-one update of $\Sigma_Q^{- 1}$, we can exploit the Sherman-Morrison
formula ({\cite{bernstein2009matrix}}, Fact 2.16.3) to compute
$\Sigma_{\ensuremath{\operatorname{RS}}}$: {\ttfamily{}}
\[ \Sigma_{\ensuremath{\operatorname{RS}}} = \Sigma_Q + \frac{\Sigma_Q
   \mathbb{G} \Sigma_Q}{1 - (\Sigma_Q)_{11} G_{11}}, \qquad
   \vec{\mu}_{\ensuremath{\operatorname{RS}}} =
   \Sigma_{\ensuremath{\operatorname{RS}}} (\Sigma_Q^{- 1} \vec{\mu}_Q +
   \vec{h} + \vec{\chi}) \]
In particular, $(\Sigma_{\ensuremath{\operatorname{RS}}})_{11}^{- 1} =
(\Sigma_Q)_{11}^{- 1} - \Delta^2 (\zeta - q)$ and
$\frac{(\mu_{\ensuremath{\operatorname{RS}}})_1}{(\Sigma_{\ensuremath{\operatorname{RS}}})_{11}}
= \frac{(\mu_Q)_1}{(\Sigma_Q)_{11}} + \sum_r \frac{(\Sigma_Q)_{1
r}}{(\Sigma_Q)_{11}} h_r + J m + \Delta \sqrt{q} t$. It follows that
$\tau_1^{- 1} = \nu_1^{- 1} - \Delta^2 (\zeta - \eta)$ and
$\frac{\alpha_1}{\tau_1} = \frac{\mu_1}{\nu_1} + \sum_r \frac{(\Sigma_Q)_{1
r}}{(\Sigma_Q)_{11}} h_r + J m + \Delta \sqrt{q} t$. In turn $\alpha_1,
\tau_1$ are enough to determine $Q_{\ensuremath{\operatorname{RS}}}^{(r)}
(v|t)$, from which the order parameters can be determined by fixed-point
interation.

Having determined the order parameters $m, q, \zeta$, we can now compute
$Q_{\ensuremath{\operatorname{RS}}}^{(r)} (v|t)$ for any $r$ and then
integrate to obtain the flux histogram $\rho_r (v) \approx \int
Q_{\ensuremath{\operatorname{RS}}}^{(r)} (v|t) \mathrm{D} t$. This integration
can be done analytically. First note that
$\vec{\mu}_{\ensuremath{\operatorname{RS}}}$ is an affine function of $t$,
explicitly $\vec{\mu}_{\ensuremath{\operatorname{RS}}} =
\vec{\mu}_{\ensuremath{\operatorname{RS}}}^{(0)} + t
\vec{\mu}_{\ensuremath{\operatorname{RS}}}^{(1)}$, where
$\vec{\mu}_{\ensuremath{\operatorname{RS}}}^{(0)} =
\Sigma_{\ensuremath{\operatorname{RS}}} [\Sigma_Q^{- 1} \vec{\mu}_Q + \vec{h}
+ \vec{\chi} (0)]$ and $\vec{\mu}_{\ensuremath{\operatorname{RS}}}^{(1)} =
\Delta \sqrt{q} \Sigma_{\ensuremath{\operatorname{RS}}} \vec{e}_1$. Similarly,
$\alpha_r = \alpha^{(0)}_r + t \alpha^{(1)}_r$, where
$\frac{\alpha_r^{(0)}}{\tau_r} =
\frac{(\mu_{\ensuremath{\operatorname{RS}}})_r^{(0)}}{(\Sigma_{\ensuremath{\operatorname{RS}}})_{r
r}} - \frac{a_r}{d_r}$ and $\frac{\alpha_r^{(1)}}{\tau_r} =
\frac{(\mu_{\ensuremath{\operatorname{RS}}})_r^{(1)}}{(\Sigma_{\ensuremath{\operatorname{RS}}})_{r
r}}$.

Since $\alpha_r$ is an affine function of $t$, we can write $\alpha_r = t
\alpha_r^{(1)} + \alpha_r^{(0)}$, where the constant $\alpha_r^{(0)}$ is the
value of $\alpha_r$ for $t = 0$ and $\alpha_r^{(1)} =
\frac{(\Sigma_{\ensuremath{\operatorname{RS}}})_{r
1}}{(\Sigma_{\ensuremath{\operatorname{RS}}})_{r r}} \tau_r \Delta \sqrt{q}$.
Then:
\begin{eqnarray*}
  \rho_r (v) & \approx & \int Q_{\ensuremath{\operatorname{RS}}}^{(r)} (v|t)
  \mathrm{D} t \propto \psi_r (v_r) \exp \left[ - \frac{1}{2} \frac{(v_r -
  \alpha_r^{(0)})^2}{(\alpha_r^{(1)})^2 + \tau_r} \right]
\end{eqnarray*}
which is a univariate normal with variance $(\alpha_r^{(1)})^2 + \tau_r$
(which can negative) and mean $\alpha_r^{(0)}$, truncated to the interval
$[\ensuremath{\operatorname{lb}}_r, \ensuremath{\operatorname{ub}}_r]$.

\section{Proof that $0 \leqslant q_r \le \zeta_r$ in the RS solution}

For any two random variables $x, y$, we have that $\langle x y \rangle \le
\max \{ \langle x^2 \rangle, \langle y^2 \rangle \}$. The proof is simple.
First note that $(x - y)^2 \ge 0$ implies $x y \le \frac{1}{2} (x^2 + y^2)$.
Therefore:
\[ \langle x y \rangle = \int \mathcal{P} (x, y) x y \mathrm{d} x \mathrm{d} y
   \le \frac{1}{2} \int \mathcal{P} (x, y) (x^2 + y^2)  \mathrm{d} x
   \mathrm{d} y = \frac{1}{2} (\langle x^2 \rangle + \langle y^2 \rangle) \le
   \max \{ \langle x^2 \rangle, \langle y^2 \rangle \} \]
where $\mathcal{P} (x, y)$ denotes an arbitrary probability distribution.
Applying this generic inequality to the variables $v_r^{\alpha}, v_r^{\beta}$,
we find that $\langle v_r^{\alpha} v_r^{\beta} \rangle \le \max \{ \langle
(v_r^{\alpha})^2 \rangle, \langle (v_r^{\beta})^2 \rangle \}$. Since under the
RS {{\em ansatz\/}} $\langle v_r^{\alpha} v_r^{\beta} \rangle = q_r$, $\langle
(v_r^{\alpha})^2 \rangle = \zeta_r$ are independent of the replica indexes
$\alpha, \beta$, it follows that $q_r \le \max \{ \zeta_r, \zeta_r \} =
\zeta_r$ as claimed. On the other hand, $q_r \geqslant 0$ is a trivial
consequence of {\eqref{saddleRS}} in the main text.

\section{Saddle-point equation for $q_r$}

Let $\mathcal{F}_{\ensuremath{\operatorname{RS}}} = \ln
\ensuremath{\operatorname{Tr}}
\mathrm{e}^{-\mathcal{H}_{\ensuremath{\operatorname{RS}}}}$. The derivation of
this equation follows:
\begin{eqnarray*}
  \frac{\Delta_r^2}{2} q_r & = & \int  \frac{\partial
  \mathcal{F}_{\ensuremath{\operatorname{RS}}}}{\partial q_r} \prod_s
  \frac{\mathrm{e}^{- \frac{t_s^2}{2}} \mathrm{d} t_s}{\sqrt{2 \pi}} = \int 
  \left( \frac{t_r}{2 q_r} \frac{\partial
  \mathcal{F}_{\ensuremath{\operatorname{RS}}}}{\partial t_r} - \frac{\partial
  \mathcal{F}_{\ensuremath{\operatorname{RS}}}}{\partial \zeta_r} \right)
  \prod_s \frac{\mathrm{e}^{- \frac{t_s^2}{2}} \mathrm{d} t_s}{\sqrt{2 \pi}}\\
  & = & \int  \left( \frac{1}{2 q_r} \frac{\partial^2
  \mathcal{F}_{\ensuremath{\operatorname{RS}}}}{\partial t_r^2} -
  \frac{\partial \mathcal{F}_{\ensuremath{\operatorname{RS}}}}{\partial
  \zeta_r} \right) \prod_s \frac{\mathrm{e}^{- \frac{t_s^2}{2}} \mathrm{d}
  t_s}{\sqrt{2 \pi}} \hspace{4em} \text{(integration by parts in the first
  term)}\\
  & = & \int  \left( \frac{\Delta_r^2}{2}
  \frac{\ensuremath{\operatorname{Tr}}
  \mathrm{e}^{-\mathcal{H}_{\ensuremath{\operatorname{RS}}}}
  v_r^2}{\ensuremath{\operatorname{Tr}}
  \mathrm{e}^{-\mathcal{H}_{\ensuremath{\operatorname{RS}}}}} -
  \frac{\Delta_r^2}{2} \left( \frac{\ensuremath{\operatorname{Tr}}
  \mathrm{e}^{-\mathcal{H}_{\ensuremath{\operatorname{RS}}}}
  v_r}{\ensuremath{\operatorname{Tr}}
  \mathrm{e}^{-\mathcal{H}_{\ensuremath{\operatorname{RS}}}}} \right)^2 -
  \frac{\Delta_r^2}{2} \frac{\ensuremath{\operatorname{Tr}}
  \mathrm{e}^{-\mathcal{H}_{\ensuremath{\operatorname{RS}}}}
  v_r^2}{\ensuremath{\operatorname{Tr}}
  \mathrm{e}^{-\mathcal{H}_{\ensuremath{\operatorname{RS}}}}} \right) \prod_s
  \frac{\mathrm{e}^{- \frac{t_s^2}{2}} \mathrm{d} t_s}{\sqrt{2 \pi}}
\end{eqnarray*}

\section{Identities involving the normal distribution}

We summarize here some standard identities involving the normal density
distribution that were used above.

We can write a multivariate normal density in two equivalent ways:
\[ \exp \left[ - \frac{1}{2} (\underline{v} - \underline{\mu})^{\top}
   \Sigma^{- 1} (\underline{v} - \underline{\mu}) \right] \propto \exp \left[
   - \frac{1}{2} \underline{v}^{\top} \Sigma^{- 1} \underline{v} +
   \underline{c}^{\top} \underline{v} \right], \qquad \Sigma^{- 1}
   \underline{\mu} = \underline{c} \]
where $\Sigma^{- 1}$ is a positive definite symmetric matrix.

The product of two multivariate normal densities is again a multivariate
normal, with:
\[ \exp \left[ - \frac{1}{2} (\underline{v} - \underline{\mu}_1)^{\top}
   \Sigma_1^{- 1} (\underline{v} - \underline{\mu}_1) \right] \exp \left[ -
   \frac{1}{2} (\underline{v} - \underline{\mu}_2)^{\top} \Sigma_2^{- 1}
   (\underline{v} - \underline{\mu}_2) \right] \propto \exp \left[ -
   \frac{1}{2} (\underline{v} - \underline{\mu})^{\top} \Sigma^{- 1}
   (\underline{v} - \underline{\mu}) \right] \]
where
\[ \Sigma^{- 1} = \Sigma_1^{- 1} + \Sigma_2^{- 1}, \qquad \langle
   \underline{v} \rangle = \Sigma (\Sigma_1^{- 1} \underline{\mu}_1 +
   \Sigma_2^{- 1} \underline{\mu}_2) \]
Similarly, we can divide two multivariate normal densities
\begin{eqnarray*}
  \frac{\exp \left[ - \frac{1}{2} (\underline{v} - \underline{\mu}_1)^{\top}
  \Sigma^{- 1}_1 (\underline{v} - \underline{\mu}_1) \right]}{\exp \left[ -
  \frac{1}{2} (\underline{v} - \underline{\mu}_2)^{\top} \Sigma_2^{- 1}
  (\underline{v} - \underline{\mu}_2) \right]} & \propto & \exp \left[ -
  \frac{1}{2} (\underline{v} - \underline{\mu})^{\top} \Sigma^{- 1}
  (\underline{v} - \underline{\mu}) \right]
\end{eqnarray*}
where
\[ \Sigma^{- 1} = \Sigma^{- 1}_1 - \Sigma_2^{- 1}, \qquad \underline{\mu} =
   \Sigma (\Sigma^{- 1}_1 \underline{\mu}_1 - \Sigma_2^{- 1}
   \underline{\mu}_2) \]

\section{Finite-dimensional results}

The results in the main text represent a structureless population of cells
with random couplings $J_{i j}$. But cells can also grow in more structured
environments, like epitelial tissues, where neighboring cells can accumulate
similar mutations affecting their metabolism, leading to short-range
correlations in their phenotypes. To analyze this scenario we performed Monte
Carlo simulations in a cellular lattice arranged on a 2-dimensional grid (one
can also consider 3-dimensions with similar results).

To keep the discussion simple, each cell is described by the simple metabolic
network described in the text, consisting of only three reactions, but we will
also allow here $J_3$ to take negative values, {\itshape{i.e.}}, byproducts of
one cell can be used to feed its neighbours. This model gives raise to
interesting forms of cooperation. Some examples are shown in Figure
\ref{states2d}.

\begin{figure}[h]
  \raisebox{0.0\height}{\includegraphics{replica_metabolism_v2-1.eps}}
  \caption{\label{states2d}Representative states of the 2-dimensional lattice
  population with negative average couplings, in different phases.}
\end{figure}

For low $\Delta$ and negative $J$, cells in the lattice arrange themselves so
that the byproduct secreted by a cell serves as nutrients for its neighbors.
Such a behavior has been found experimentally for lactate in cancer or muscle
cells during exercise, and is known as the lactate shuttle
{\cite{cossio2017micro,brooks1986lactate,pellerin1998evidence,brooks2000intra}}.
In analogy to spin glasses this can be called an anti-ferromagnetic phase.
However increasing $\Delta$ leads to the gradual replacement of this ordered
state with a `glassy` phase, where the overall population might exhibit
non-zero fluxes of $v_3$, which depends on the particular instantation of the
couplings. In Figure \ref{af} we show how the ordered state is gradually
replaced by a state where, in average, there is no cooperation between the
sub-lattices. This transition is smooth, in concordance with the absence of a
discontinuous phase transition in disordered 2-dimensionsional spin glasses
{\cite{mezard1987spinglass}}.

\begin{figure}[h]
  \raisebox{0.0\height}{\includegraphics{replica_metabolism_v2-2.eps}}
  \caption{\label{af}{\bfseries{(a)}} Average $v_3$ flux in each sub-lattice
  as a function of the fluctuations in the couplings. {\bfseries{(b)}}
  Fluctuations in the population average of $v_3$ in a sub-lattice, as a
  function of the fluctuations in the couplings. In both panels we employed a
  $64 \times 64$ two-dimensional lattice, setting $J = - 10$ for the secretion
  flux.}
\end{figure}